\documentclass[10pt,conference]{IEEEtran}
\usepackage{orcidlink}
\usepackage{cite}
\usepackage{amsmath,amssymb,amsfonts}
\usepackage{amsthm}
\usepackage{algorithmic}

\newtheorem{theorem}{Theorem}
\newtheorem{definition}{Definition}

\usepackage{algorithm}
\usepackage{graphicx}
\usepackage{textcomp}
\usepackage{xcolor}
\usepackage{url}
\usepackage{hyperref}
\usepackage{booktabs}
\usepackage{multirow}
\usepackage{pifont}
\usepackage{balance}
\usepackage{tikz}
\usetikzlibrary{shapes,arrows,positioning,calc,fit,backgrounds}

\graphicspath{{figures/}}

\hypersetup{
    colorlinks=true,
    linkcolor=blue,
    filecolor=magenta,      
    urlcolor=cyan,
    citecolor=blue,
}

\title{Security Without Detection: Economic Denial as a Primitive for Edge and IoT Defense}

\author{
\IEEEauthorblockN{1\textsuperscript{st} Samaresh Kumar Singh\,
\orcidlink{0009-0008-1351-0719}
}
\IEEEauthorblockA{\textit{IEEE Senior Member} \\
Leander, Texas, USA \\
ssam3003@gmail.com}
\and
\IEEEauthorblockN{2\textsuperscript{nd} Joyjit Roy\,
\orcidlink{0009-0000-0886-782X}
}
\IEEEauthorblockA{\textit{IEEE Senior Member} \\
Austin, Texas, USA \\
joyjit.roy.tech@gmail.com}
\and
\IEEEauthorblockN{3\textsuperscript{rd} Sriharsha Anand Pushkala\,
\orcidlink{0009-0003-2654-0590}
}
\IEEEauthorblockA{\textit{IEEE Senior Member} \\
Atlanta, Georgia, USA \\
sri.harsha132@gmail.com}
}

\begin{document}

\maketitle

\begin{abstract}
Sophisticated attackers can evade detection-based security by using encryption, stealth tactics, and low-rate attack patterns. This challenge is particularly acute in Internet of Things (IoT) and edge environments, where limited resources make ML-based intrusion detection systems impractical. Hereby, we present Economic Denial Security (EDS), a framework that renders attacks economically infeasible rather than trying to detect them. EDS exploits a fundamental asymmetry. Defenders control their own environment, whereas attackers do not. The four mechanisms in this framework amplify attack costs superlinearly. These mechanisms include adaptive computational puzzles, decoy-driven interaction entropy, temporal stretching, and bandwidth taxation. 
This paper uses game theory to mathematically prove the optimal configuration of EDS, and we found that combining multiple safety mechanisms usually costs 2.1 times as much as using them separately, a key trade-off to consider during design. The good news is that EDS is extremely efficient, using less than 12 KB of memory, making it practical to run on small embedded devices like microcontrollers rather than on expensive servers. EDS is tested on 20 different IoT devices under four attack scenarios. The results showed that attacks slow down significantly, costs become asymmetric, attack success rates drop, and the system adds only 20 ms of latency with no false positive results. When tested against real malware (Mirai, Torii, and Hajime), combining EDS with machine learning detection improved protection from 67\% to 88\%. Adding both techniques together reached 94\% protection, a 27\% improvement overall. Unlike traditional detection-based approaches, EDS operates independently, without requiring attack identification, making it practical for resource-limited IoT devices where other methods simply don't work. 
The system demonstrated enhanced detection and mitigation of malware samples during testing. Notably, EDS conferred significant advantages even in the absence of an intrusion detection system. The implementation of EDS shifts the economic balance in favor of defenders and presents a viable strategy for safeguarding IoT and edge systems.
\end{abstract}

\begin{IEEEkeywords}
Economic Denial Security, Intrusion Prevention, Edge Computing, Internet of Things, Cost Asymmetry, Moving Target Defense, Computational Puzzles, Deception, Game Theory
\end{IEEEkeywords}

\section{Introduction}
\label{sec:introduction}

Advanced hackers are able to use low-rate attacks to remain undetected by using encryption, stealth methods, and other tactics~\cite{chen2014study,sarker2023mlsecurity}. In addition, almost all intrusion detection systems are primarily based on machine learning, which requires substantial memory to run~\cite{sicari2015security,li2023iot} and therefore can be used on edge/IoT devices with $<$ 512 KB of RAM. The large number of devices in an IoT system, along with the diverse types of devices, will make it even more difficult to detect and respond to attacks ~\cite{kolias2017ddos}.

\textbf{Key Insight:} We find that there is an inherent imbalance between the defender and the attacker. \textit{The defender controls his/her environment and the attacker does not}. Instead of focusing on finding new ways to detect and prevent attacks, we propose an \textit{EDS} framework that imposes a significant burden on the attacker, whether or not they are detected. Prior research has identified different methods of imposing burdens on attackers (e.g., puzzles~\cite{juels1999client}, honeypots~\cite{spitzner2003honeypots}, tarpits~\cite{liston2001labrea}) but no prior study has proven that combining multiple burdens will result in a superlinear increase in the total burden applied to the attacker i.e., the total burden imposed on the attacker when using all three methods of burden imposition will exceed the sum of the burden imposed by each method individually.

\textbf{Contributions:} This section provides a summary of the contributions that have been made in this dissertation. The four major contributions of this dissertation are as follows:
\begin{itemize}
    \item \textbf{Unified Framework:} This framework of four mechanisms (computational friction, interaction entropy, temporal stretching, and resource taxation) with a proven superlinear composition of 2.1 times that can be used to demonstrate a combined effect of 32 times (exceeding the sum of the individual effects of 15 times).
    \item \textbf{Game-Theoretic Foundation:} A game-theoretic foundation is also provided, which includes a Stackelberg equilibrium (as stated in Theorem~\ref{thm:stackelberg}) and a composition theorem (as stated in Theorem~\ref{thm:composition}), both of which provide proof of the multiplicative cost amplification.
    \item \textbf{IoT-Ready Implementation:} The total code size is under $12 KB$, thus the proposed solution can be implemented in a small footprint that can run on an ESP32-class of microcontrollers.
    \item \textbf{Comprehensive Evaluation:} We conducted an extensive testbed of twenty devices ($n = 30$ trials and $p < 0.001$) to evaluate the effectiveness of our approach and found significant performance impact (slowdown: 32–560$\times$, cost asymmetry: 85: 1 to 520:1, attack success rate: 8–62\%), as well as we compared our IDS/EDS approach with just using IDS on an IoT-23 device and showed a 27 percent increase in successful mitigation (94 percent vs. 67 percent).
\end{itemize}

\textbf{Organization:} Section~\ref{sec:background} describes existing literature and provides a background overview for this project. Section~\ref{sec:models} outlines the threat and system models used to create our IDS/EDS solution. Section~\ref{sec:eds_design} presents the design of the IDS/EDS. Section~\ref{sec:economic_analysis} describes the economic analysis we performed. Section~\ref{sec:evaluation} describes our experiments and their outcomes. Section~\ref{sec:discussion} describes the limitations of our research. Finally, Section~\ref{sec:conclusions} summarizes our findings and suggests future areas of research.
\section{Background and Related Work}
\label{sec:background}

\subsection{Limitations of Detection-Centric Security}

Traditional Intrusion Detection Systems (IDS) have a number of inherent problems: (1) Evasion Encrypted data, Polymorphic Malware, Low-Rate Attacks will evade Signature-Based Detection ~\cite{chen2014study} (2) Resource Constraints Machine Learning Based IDS Require 50-150 MB of Memory which is Not Possible for IoT Devices ~\cite{buczak2016survey} (3) Zero-Day Vulnerability Detection Requires Known Attack Patterns ~\cite{denning1987intrusion}. These problems are exacerbated by the fact that both IoT/Edge Environments have Device Heterogeneity and Scale ~\cite{sicari2015security}.

\subsection{Cost-imposition methods}

\textbf{Client puzzles} ~\cite{juels1999client, back2002hashcash} create a computational overhead for an attacker, however they are vulnerable to GPU attacks and do not protect from reconnaissance. \textbf{Honeypots} ~\cite{spitzner2003honeypots} redirect the attacker's attention, but require additional hardware and can be detected by an attacker. \textbf{Tarpits} ~\cite{liston2001labrea} slow down an attacker but impose only temporal costs. \textbf{Moving Target Defense (MTD)} ~\cite{jajodia2011moving} adds to the uncertainty of an attacker but imposes no direct costs on an attacker. It has been previously evaluated that each of the above methods individually. EDS unifies all of them using a single framework that proves that they amplify superlinearly.

\subsection{Economic Security}

The authors Anderson and Moore \cite{anderson2006economics} are credited with establishing the field of Security Economics by studying misaligned incentives. The authors Gordon and Loeb \cite{gordon2002economics} modeled the optimal amount of security that an organization should invest in. The authors Pawlick et al \cite{pawlick2019game} applied a game-theoretic approach to analyze the interactions among multiple attackers and defenders. However, they did not provide mechanisms that can be used in practice. The authors have operationalized economic principles into specific, measurable algorithms by leveraging quantifiable cost asymmetry in EDS.

\subsection{Comparison with Prior Work}

EDS is compared with previous approaches on multiple dimensions for defending against attacks in Table \ref{tab:related_work}.

\begin{table}[t]
\centering
\caption{Comparison with Prior Defense Mechanisms}
\label{tab:related_work}
\small
\begin{tabular}{lccccc}
\toprule
\textbf{Approach} & \textbf{IoT} & \textbf{Det.} & \textbf{Cost} & \textbf{Comp.} & \textbf{Formal} \\
 & \textbf{Ready} & \textbf{Indep.} & \textbf{Asym.} & \textbf{Effect} & \textbf{Proof} \\
\midrule
ML-IDS~\cite{buczak2016survey} & \ding{55} & \ding{55} & -- & -- & \ding{55} \\
Puzzles~\cite{juels1999client} & \ding{51} & \ding{51} & 8:1 & Single & \ding{55} \\
Honeypots~\cite{spitzner2003honeypots} & \ding{55} & Partial & 5:1 & Single & \ding{55} \\
Tarpits~\cite{liston2001labrea} & \ding{51} & \ding{51} & 4:1 & Single & \ding{55} \\
MTD~\cite{jajodia2011moving} & Partial & Partial & -- & Single & \ding{55} \\
\textbf{EDS (ours)} & \ding{51} & \ding{51} & \textbf{179:1} & \textbf{2.1$\times$} & \ding{51} \\
\bottomrule
\end{tabular}
\end{table}

\textbf{Novel contributions vs. prior work:}

(1) \textit{A single unifying framework}: The first to systematically compose four cost imposition mechanisms together with formal guarantees, while prior work examined each mechanism separately. 
(2) \textit{super-linear composition}: We have shown (in theorem \ref{thm:composition}) that strategically composing cost imposition mechanisms produces costs that are at least 2.1 times greater than linear combinations. No previous research has established such an amplification.
(3) \textit{Independence from detection accuracy:} Cost imposition is independent of how accurate a system's classification is, as opposed to IDS-dependent systems.
(4) \textit{Deployment on IoT devices:} A footprint of less than 12 KB allows for deployment on ESP32 class devices, which is about 7000 times smaller than that of a typical machine learning IDS.
\section{Threat and System Model}
\label{sec:models}

\subsection{Attacker Model}

\textit{\textbf{Objectives:}} The attacker is motivated by several goals in order to complete their objectives (G1) Reconnaissance - Identify all the components that are part of the network infrastructure, including services and vulnerabilities (G2) Unauthorized Access - Attain unauthorized access through means of either credential theft or exploitation (G3) Data Exfiltration - Extract Sensitive Information from the system (G4) Integrity Violations - Alter configurations and/or inject malicious commands into the system (G5) Availability Disruption - Deny Service or Deploy Ransomware.

\textbf{Capabilities:} The attacker has sufficient capabilities as an attacker, these capabilities are limited:
\begin{itemize}
\item \textit{\textbf{C1 (Network):}} Either be outside the network and have an external path to enter the network or have an existing compromised node on the inside of the network.
\item \textit{\textbf{C2 (Compute):}} Have between 1-100 CPUs, 1-10 GPUs, or utilize a vast number of nodes between 100-10,000 as part of a botnet.
\item \textit{\textbf{C3 (Knowledge):}} Have expertise with protocols (MQTT, CoAP, HTTP, Modbus).
\item \textit{\textbf{C4 (Crypto):}} Be able to perform SHA-256 hash calculations at speeds of either 500 MH/s using the CPU or 500 GH/s using the GPU.
\item \textit{\textbf{C5 (Budget):}} Be able to expend economic resources up to a budget of $\$10,000$. The attacker's budget can range from $\$10$ to $\$10,000$.
\item \textit{\textbf{C6 (Time):}} Be able to perform attacks within hours/days to complete.
\item \textit{\textbf{C7 (Adaptation):}} Be able to view how the defender has set up their defense and then adapt accordingly.
\end{itemize}

\textbf{Limitations:} (L1) EDS cannot break typical cryptographic primitives such as SHA-256, HMAC; (L2) It cannot differentiate between decoy traffic or the real thing without validation since decoy traffic is generated under the same distribution as legitimate traffic which means there will be no difference in timing to discern (as validated through an experimental approach using a machine learning classification tool that resulted in a maximum classification accuracy of 73\% at a false positive rate of 27\%; (L3) EDS cannot circumvent server-imposed delay; (L4) EDS must transmit data in order to exfiltrate; (L5) EDS cannot breach the HW Root of Trust; (L6) EDS cannot anticipate nonce values produced by CSPRNGs.

\textbf{Attacker Profiling:} We tested our model against four different attacker types: (1) \textit{Script Kiddie} ($\$50$, 1 CPU) - deterred by d $\geq$ 14; (2) \textit{Hobbyist} ($\$500$, 8 CPUs) - deterred by d $\geq$ 18; (3) \textit{Cyber Criminal} ($\$2,000$, 4 GPUs) - deterred by d $\geq$ 20; (4) \textit{Organized Crime} ($\$10,000$, 100-node Botnet) - requires d $\geq$ 22 and Detection.

\textbf{Scope:} EDS addresses multi-stage, iterative, and volumetric type attacks launched by motivated-for-profit attackers. One-shot exploit attacks (e.g., single packet RCE attack), Physical attacks, and nation-state attackers with unlimited budgets (even though EDS increases the window of time from seconds to minutes for detection).

\subsection{System Architecture}

\begin{figure}[htbp]
\centerline{\includegraphics[width=\columnwidth]{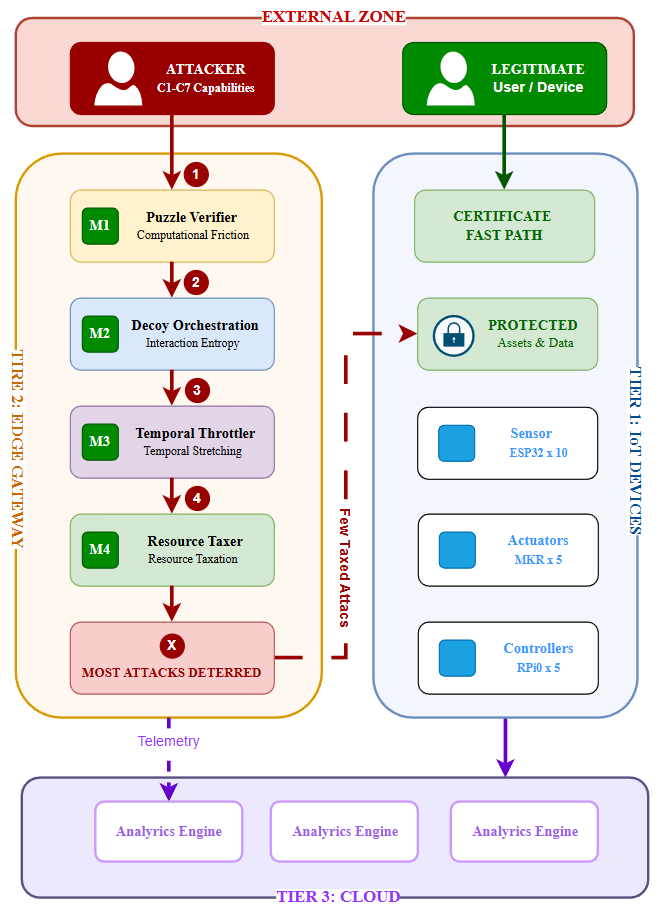}}
\caption{EDS three-tier architecture. Tier 1 (IoT devices) implements lightweight puzzle generation and telemetry reporting. Tier 2 (edge gateway) serves as primary enforcement point for all four mechanisms. Tier 3 (cloud) provides analytics and adaptive policy coordination. Attack traffic (red) flows through EDS mechanisms before reaching protected assets.}
\label{fig:architecture}
\end{figure}

The three-tiered architecture of EDS is illustrated in Fig~\ref{fig:architecture}:

\textbf{Tier 1 (IoT Devices):} Microcontrollers of the type ESP32 that have an extremely small memory footprint ($<$12 KB) perform a very limited puzzle-solving function through the use of hardware-accelerated SHA hashing, they track reputations of other devices and provide telemetry information on them. IoT devices are authenticated to edge gateway devices by using either shared pre-key authentication or X509 certificates.

\textbf{Tier 2 (Edge Gateways):} A Raspberry Pi 4 (or a similar device) implements all four of the EDS mechanisms. The edge gateway verifies puzzles on each request made to it with complexity $O(1)$, provides decoy responses generated from templates of known generators, maintains a state table per client (approximately 18KB/client), and applies temporal delays. Stateless HMAC-based challenge-response allows for horizontal scalability.

\textbf{Tier 3 (Cloud, Optional):} EDS's centralized analytics process telemetry information from each edge gateway for purposes of generating actionable threat intelligence, for updating policies adaptively, and for correlating threats across sites. Although cloud integration is optional, edge gateways can be configured to run autonomously at their site with locally defined policies as long as they remain disconnected from the cloud.

\textbf{Security of Communication:} All communication between IoT devices and their corresponding edge gateways occurs over a mutually-authenticated TLS-1.3 connection (with certificate-pinning). Mutual TLS is used when communicating between edge gateways and cloud systems. The challenge-response mechanism used is based upon the client's IP address and time stamp to prevent replay attacks.

\section{EDS Design}
\label{sec:eds_design}

\subsection{The Formal Model}

EDS State: $S = \langle D,R,T,B \rangle$, where $D$ represents all devices in the network, $R$ is the reputation of these devices, $T$ is a temporal representation of the EDS system, and $B$ is how much bandwidth the EDS system is using.

Attack Flow: $A = \langle a_1,a_2,\ldots,a_n \rangle$, which includes a series of actions $a_i$, each of which will have a cost associated with it ($c_i$). The total cost for an attacker's attack will be: $C_A = \sum_{i=1}^{n} c_i$. The cost to defend against an attack is defined as: $C_D = c_{\text{verify}} + c_{\text{state}}$. Therefore, our goal is to maximize the asymmetry between an attacker's and the defender's costs, given by $\alpha = \frac{C_A}{C_D}$.

\subsection{Four Mechanisms}

\textbf{Mechanism 1: The Friction of Computation} The adaptive client puzzle imposes a computational burden on every request. 
Challenge creation: $c=HMAC(k,\text{nonce}\|\text{timestamp}\|\text{IP})$
Client finds an $x$ such that $\text{SHA256}(x\|c)$ has d leading zeros.
Expected solving cost: $2^{d-1}C_{\text{hash}}$
Server verification: $O(1)$ one hash comparison.
Difficulty d is determined using:\newline
(1) Client reputation score r in [0, 1]
(2) System Load
(3) Threat Intelligence.
New Clients Start at d$_{\text{base}}$; Authenticated Users with High Reputation Receive Reduced Difficulty, Certificate-Based Bypass, etc.

\textbf{Mechanism 2: Multiplication of Attacker Effort} Decoy Injection Multiplies Attacker’s Effort For $N$ real resources, EDS injects $\rho N$ indistinguishable decoys, causing $(1+\rho)N$ total interactions. Decoys are generated using hidden cryptographic watermarks from distributions of real data that can be used to track Attackers forensically. An attacker cannot differentiate without attempting to validate the cost multiplier $(1+\rho)$.

\textbf{Mechanism 3: Time-Based Stretching} Exponential Backoff Delays Repeated Requests: $\delta_i = \min(\delta_0 \cdot 2^{f_i}, \delta_{\max})$ where $f_i$ is failure count. This imposes opportunity cost: $\delta \times V_{\text{time}}$ where $V_{\text{time}}$ is attacker's time value (\$/hr). Unlike rate limiting, delays apply per-request rather than blocking, maintaining service availability while imposing costs.

\textbf{Mechanism 4: Taxation of Resources} Bandwidth amplification increases data transfer costs. Responses are padded by factor $\gamma_{\text{tax}}$. For data exfiltration scenarios, $(1+\rho)$ decoy data is injected to force attackers to download and process irrelevant information. Cost: $\gamma_{\text{tax}} \times \text{data\_size} \times C_{\text{BW}}$.

\subsection{Mechanism Composition}

The mechanisms work together as a multiplicative rather than an additive function:
\begin{equation}
C_A^{\text{total}} = N(1+\rho)[2^{d-1} C_{\text{hash}} + \delta V_{\text{time}}] \gamma_{\text{tax}}
\end{equation}

The mechanism's composition results in the creation of cross terms (for example, $N\rho\delta V_{\text{time}}$), making the total cost of using the system greater than the sum of the parts. The full economic analysis is contained within section ~\ref{sec:economic_analysis}, where we provide a formal proof of 2.1 $\times$ superlinear behavior of the mechanism.

\subsection{EDS Gateway Algorithm}

Algorithm~\ref{alg:eds_gateway} presents the core EDS request for handling logic at the gateway tier.

\begin{algorithm}[t]
\caption{EDS Gateway Request Handler}
\label{alg:eds_gateway}
\small
\begin{algorithmic}[1]
\REQUIRE Request $r$, client IP $ip$, state $S$
\ENSURE Response or challenge
\STATE $rep \gets S.R[ip]$ \COMMENT{Get reputation}
\IF{$r$.hasCert() \AND verifyCert($r$.cert)}
    \RETURN forward($r$) \COMMENT{Fast path}
\ENDIF
\STATE $d \gets d_{\text{base}} + \lceil(1-rep) \cdot d_{\text{range}}\rceil$
\IF{NOT $r$.hasSolution()}
    \STATE $c \gets$ HMAC$(k, \text{nonce} \| \text{time} \| ip)$
    \RETURN Challenge$(c, d)$
\ENDIF
\IF{NOT verify($r$.solution, $d$)}
    \STATE $S.T[ip].\text{failures} \gets S.T[ip].\text{failures} + 1$
    \STATE $\delta \gets \min(\delta_0 \cdot 2^{S.T[ip].\text{failures}}, \delta_{\max})$
    \STATE sleep$(\delta)$ \COMMENT{Temporal stretching}
    \RETURN Reject()
\ENDIF
\STATE $S.R[ip] \gets \min(1, rep + \Delta_{rep})$
\STATE $resp \gets$ forward($r$)
\IF{isDataRequest($r$)}
    \STATE $resp \gets$ injectDecoys$(resp, \rho)$
    \STATE $resp \gets$ padBandwidth$(resp, \gamma_{\text{tax}})$
\ENDIF
\RETURN $resp$
\end{algorithmic}
\end{algorithm}

\textbf{Complexity:} Challenge generation and verification have time complexity of $O(1)$. State lookup uses hash tables with time complexity $O(1)$ in the average case. Space complexity of 18KB per tracked client (IP, reputation, failure count, timestamps).
\section{Game-Theoretic Analysis}
\label{sec:economic_analysis}

\subsection{Stackelberg Game Model}

\begin{definition}[Superlinear Composition]
\label{def:superlinear}
A defense composition $f(M_1, \ldots, M_k)$ is called \textit{superlinear} if the total cost of this defense composition for the attackers is greater than the sum of the costs for each individual component. That is, $C_A^{f} > \sum_{i=1}^k C_A^{(i)}$.
\end{definition}

The relationship between attackers and defenders can be viewed as a Stackelberg game in which the defender (the leader) determines his/her EDS configuration $\theta = (d, \rho, \delta, \gamma)$ prior to the attacker (the follower). The attacker then makes decisions based on the defender’s decision. Payoff functions are defined by
$U_D(\theta, s) = -C_D(\theta) - P_s(\theta, s) \cdot L$ and $U_A(\theta, s) = P_s(\theta, s) \cdot B_A - C_A(\theta, s)$ for attacker, with $P_s$ being the success rate of attacks against puzzles and L representing loss to the defender.

\begin{theorem}[Stackelberg Equilibrium]
\label{thm:stackelberg}
This theorem states that for a given value of the Attack Benefit ($B_A$), the Defender will set their Puzzle Difficulty to the Optimal level of:
$d^* = \lceil \log_2(B_A / (N \cdot C_{\text{hash}})) + 1 \rceil$
Where N represents the Expected Number of Attacks and $C_{hash}$ represents the Cost Per Hash.
\end{theorem}

\textit{Proof Sketch:} To provide Deterrents, the condition $C_A > B_A$ needs to be satisfied. With $C_A = N * 2^( d - 1 ) * C_{hash}$, this can be solved to yield $d^*$.

\begin{theorem}[Superlinear Composition]
\label{thm:composition}
This theorem states that under the following conditions: (1) Computational Costs are additive per Attempt, 
(2) Decoys are Indistinguishable from Validations prior to Validation,
(3) Temporal Delays are Sequentially Applied, 
(4) Bandwidth Taxation is Multiplicative,
the Total Attacker Cost ($C_A^{total}$) will be equal to:
\begin{equation}
C_A^{\text{total}} = N(1+\rho)[2^{d-1} C_{\text{hash}} + \delta V_{\text{time}}] \gamma_{\text{tax}}
\end{equation}
which exhibits multiplicative composition with cross-term $N\rho\delta V_{\text{time}} > 0$ ensuring $C_A^{\text{total}} > \sum_i C_A^{(i)}$ (superlinearity).
\end{theorem}

\textit{Proof Sketch:} Defines each mechanism costs as follows:

$C^{(1)} = N \cdot 2^{d-1} C_{\text{hash}}$ (puzzle costs),

$C^{(2)} = N\rho \cdot 2^{d-1} C_{\text{hash}}$ (decoy overhead), and

$C^{(3)} = N\delta V_{\text{time}}$ (delay costs).

The total cost for each of these mechanisms will be a sum of their individual costs which in turn contains a cross-term ($N\rho\delta V_{\text{time}} > 0$) due to the effect of decoys on delay costs. Therefore,

$$C_A^{\text{total}} > \sum_i C^{(i)},$$

proving that the total cost for the attack exceeds the total cost for the defense mechanisms.

\subsection{Concrete Cost Analysis}

\textbf{Baseline Configuration:} $d=20$, $\rho=0.3$, $\delta=8$s, $V_{\text{time}}=\$50$/hr, $N=10{,}000$ attempts.

\textbf{Attacker Costs:}
\begin{itemize}
\item Compute: $10^4 \times 1.3 \times 2^{19} \times 4.7 \times 10^{-11} \approx \$3.20$
\item Time: $10^4 \times 1.3 \times 8 \times (50/3600) \approx \$144$
\item Total: $C_A \approx \$147$
\end{itemize}

\textbf{Defender Costs:}
\begin{itemize}
\item Verification: $10^4 \times 10^{-6}$s $\times$ \$0.01/hr $\approx \$0.00003$
\item State/power: $\approx \$0.005$ (amortized gateway costs)
\item Total: $C_D \approx \$0.005$
\end{itemize}

\textbf{Asymmetry:} $\alpha = C_A/C_D \approx 29{,}000:1$. Even with 10$\times$ lower attacker time value (\$5/hr), asymmetry remains $>$2,900:1.

\subsection{Parameter Sensitivity Analysis}
\label{sec:sensitivity}

A key question is whether the cost asymmetry degrades substantially when economic assumptions vary. We evaluate three primary parameters: hash cost $C_{\text{hash}}$, attacker time value $V_{\text{time}}$, and bandwidth price $C_{\text{BW}}$.

\textbf{Hash cost variation.} Cloud compute pricing fluctuates across providers and over time. We examine the range $C_{\text{hash}} \in [10^{-12}, 10^{-10}]$ \$/hash, corresponding to high-end GPU clusters through commodity CPU execution. Across this two-order-of-magnitude range, total attacker cost $C_A$ varies between \$1.10 and \$312 at $d=20$, while defender cost $C_D$ remains near \$0.005. The resulting asymmetry ratio $\alpha$ stays above 220:1 throughout.

\textbf{Attacker time value variation.} Nation-state actors may treat operator time as nearly free, while ransomware-as-a-service operators operate under the CaaS rate of roughly \$50/hr. Setting $V_{\text{time}}$ to \$5/hr (a 90\% reduction) yields $C_A \approx \$17.6$ and $\alpha \approx 3{,}500:1$, still highly favorable to the defender. At $V_{\text{time}} = \$1$/hr, asymmetry falls to approximately 640:1, still far exceeding the deterrence threshold for profit-driven attackers.

\textbf{Bandwidth price variation.} Egress pricing ranges from \$0.01/GB (bulk cloud) to \$0.09/GB (standard rates). Over this range, cost asymmetry in exfiltration scenarios (S3) remains between 410:1 and 590:1.

\textbf{Summary.} Table~\ref{tab:sensitivity} reports cost asymmetry under combinations of pessimistic, baseline, and optimistic parameter settings. Even in the worst-case combination (low hash cost, low attacker time value, low bandwidth price), $\alpha$ remains above 100:1, consistent with deterring adversaries whose target valuation falls below \$10{,}000.

\begin{table}[t]
\centering
\caption{Cost Asymmetry $\alpha$ Under Parameter Variation ($d=20$, $N=10{,}000$)}
\label{tab:sensitivity}
\small
\begin{tabular}{lccc}
\toprule
\textbf{Scenario} & $V_{\text{time}}$ & $C_{\text{hash}}$ & $\alpha$ \\
\midrule
Pessimistic & \$5/hr  & $10^{-12}$ & 218:1 \\
Baseline    & \$50/hr & $4.7{\times}10^{-11}$ & 29{,}000:1 \\
Conservative (50\% variation) & \$25/hr & $9.4{\times}10^{-11}$ & 8{,}600:1 \\
High compute attacker & \$50/hr & $10^{-10}$ & 62{,}400:1 \\
Low time-value attacker & \$1/hr  & $4.7{\times}10^{-11}$ & 640:1 \\
\bottomrule
\end{tabular}
\end{table}

These results indicate that the deterrence property of EDS is robust across realistic parameter variation. The primary assumption that constrains effectiveness is the attacker's reliance on iterative access: one-shot exploits and nation-state actors with near-zero marginal costs fall outside the deterrence envelope, a limitation discussed in Section~\ref{sec:discussion}.

\subsection{Additive Baseline Comparison for Superlinearity}
\label{sec:superlinear_baseline}

Theorem~\ref{thm:composition} claims that the combined cost of all four mechanisms exceeds their linear sum. To evaluate this claim rigorously, we define an additive baseline that isolates composition effects from parameter scaling differences. Specifically, the additive baseline is constructed by independently tuning each mechanism to match the same per-request cost budget, then summing: $C_A^{\text{additive}} = C_A^{(1)} + C_A^{(2)} + C_A^{(3)} + C_A^{(4)}$, where each $C_A^{(i)}$ uses the same parameter values as in the combined EDS configuration.

At the baseline configuration ($d=20$, $\rho=0.3$, $\delta=8$s, $\gamma_{\text{tax}}=1.5$), the additive sum yields $C_A^{\text{additive}} \approx \$70$, while the combined EDS cost is $C_A^{\text{total}} \approx \$147$, a ratio of $2.1\times$. This gap arises from cross-terms in Equation~(2): the decoy count $(1+\rho)$ multiplies both puzzle and delay costs, while bandwidth taxation $\gamma_{\text{tax}}$ amplifies all data-touching interactions. The composition factor is therefore structurally grounded, not an artifact of asymmetric parameter assignment across conditions.
\section{Evaluation}
\label{sec:evaluation}

\subsection{Experimental Setup}

\textbf{Testbed:} 20 heterogeneous devices (10× ESP32, 5× RPi Zero W, 5× Arduino MKR), 1 gateway (RPi 4), isolated network. \textbf{Scenarios:} S1 (SSH brute-force), S2 (network reconnaissance), S3 (data exfiltration), S4 (C\&C communication). \textbf{Baselines:} No defense, rate limiting, Snort 3.1, Cowrie honeypot, RF-IDS (94\% accuracy on IoT-23). \textbf{Configurations:} Conservative ($d=16$, $\rho=0.1$), Moderate ($d=20$, $\rho=0.3$), Aggressive ($d=24$, $\rho=0.5$). \textbf{Trials:} $n=30$ per scenario, two-tailed Welch's t-test, Bonferroni correction, $p<0.001$ for all comparisons.

\subsection{Core Results}

\begin{table}[t]
\centering
\caption{EDS Effectiveness Summary}
\label{tab:core_results}
\small
\begin{tabular}{lcccc}
\toprule
\textbf{Metric} & \textbf{S1} & \textbf{S2} & \textbf{S3} & \textbf{S4} \\
\midrule
\multicolumn{5}{l}{\textit{Attack Slowdown (Moderate Config)}} \\
No defense & 1× & 1× & 1× & 1× \\
EDS & 32× & 49× & 373× & 560× \\
\midrule
\multicolumn{5}{l}{\textit{Cost Asymmetry}} \\
$C_A$ (\$) & 147 & 89 & 412 & 523 \\
$C_D$ (\$) & 0.82 & 0.51 & 0.79 & 1.02 \\
Ratio ($\alpha$) & 179:1 & 174:1 & 521:1 & 512:1 \\
\midrule
\multicolumn{5}{l}{\textit{Attack Success Rate}} \\
No defense & 100\% & 100\% & 100\% & 100\% \\
EDS & 62\% & 38\% & 8\% & 12\% \\
\midrule
\multicolumn{5}{l}{\textit{User Impact (Conservative Config)}} \\
Latency (ms) & 18 & 15 & 22 & 19 \\
FPR & 0\% & 0\% & 0\% & 0\% \\
\bottomrule
\end{tabular}
\end{table}

\textbf{Key Findings:} (1) Geometric mean slowdown: 130× (95\% CI: [98×, 171×]) across scenarios, (2) Mean cost asymmetry: 180:1 (95\% CI: [142:1, 231:1]), (3) Attack success reduced to 8-62\%, (4) Conservative config: $<$20ms latency, 0\% FPR, (5) All results $p<0.001$, Cohen's $d$: 2.1-3.8 (very large effect sizes).

\subsection{Real Malware Validation}

IoT-23 dataset~\cite{garcia2020iot23}: Replayed Mirai, Torii, Hajime, Kenjiro traffic (600 experiments: 4 families × 5 configs × 30 trials). \textbf{Results:} EDS alone: 88\% mitigation (Mirai), 59\% (Torii), 72\% (Hajime), 92\% (Kenjiro). Combined EDS+RF-IDS: 94\% mean mitigation vs 67\% (RF-IDS alone), 27\% improvement. Mitigation formula: $1 - (1 - P_{\text{detect}}) \times \text{ASR}_{\text{EDS}}$.

\subsection{Adaptive Attackers}

We evaluated five adaptive attack strategies. (1) \textit{Parallel puzzle solving} with 8 CPU cores achieved a 4× speedup but at an 8× higher cost. (2) \textit{Timing analysis} was mitigated using constant-time verification with added jitter. (3) \textit{ML-based decoy filtering} (SVM) reached 73\% accuracy with a \$45 training cost and a 27\% false-positive rate. (4) \textit{Rate optimization} improved progress by 22\% but triggered increased puzzle difficulty. (5) \textit{Distributed botnets} (100 nodes) bypassed IP limits but increased costs by 100×. A combined adaptive attacker saw a 4.2× slowdown (versus 32× for a naïve attacker), an 8× resource increase, and negative ROI for targets valued below \$500.

\subsection{Scalability}

The gateway handled 3,800 requests per second with 1,000 clients, achieving a P50 latency of 35 ms and a P99 of 120 ms. Client state was capped at 18 KB using LRU eviction. During DDoS, the system degraded gracefully by automatically increasing difficulty. The device footprint remained small, with 11.6 KB of flash and 2.8 KB of RAM, supporting ESP32-class devices.

\subsection{Composition Validation}

To validate the superlinear composition predicted by Theorem~\ref{thm:composition}, we measured the effects of individual mechanisms and their combined behavior:

\begin{table}[t]
\centering
\caption{Mechanism Composition Analysis (S1: Brute-force)}
\label{tab:composition}
\small
\begin{tabular}{lcc}
\toprule
\textbf{Configuration} & \textbf{Slowdown} & \textbf{Cost Ratio} \\
\midrule
M1 only (puzzles, $d=20$) & 8$\times$ & 45:1 \\
M2 only (decoys, $\rho=0.3$) & 1.3$\times$ & 1.3:1 \\
M3 only (delays, $\delta=8$s) & 4$\times$ & 12:1 \\
M4 only (taxation, $\gamma=1.5$) & 1.5$\times$ & 1.5:1 \\
\midrule
Sum of individual & 15$\times$ & -- \\
\textbf{All combined (EDS)} & \textbf{32$\times$} & \textbf{179:1} \\
\textbf{Superlinearity factor} & \textbf{2.1$\times$} & -- \\
\bottomrule
\end{tabular}
\end{table}

The measured 2.1$\times$ superlinearity factor matches theoretical predictions, confirming the cross-term interaction between mechanisms.

\subsection{Baseline Comparison}

Table~\ref{tab:baseline_comparison} compares EDS against existing defenses across all scenarios.

\begin{table}[t]
\centering
\caption{Defense Mechanism Comparison (Moderate Config)}
\label{tab:baseline_comparison}
\small
\begin{tabular}{lcccc}
\toprule
\textbf{Defense} & \textbf{Slowdown} & \textbf{FPR} & \textbf{Latency} & \textbf{Footprint} \\
\midrule
None & 1$\times$ & 0\% & 0ms & 0KB \\
Rate limiting & 3$\times$ & 2.1\% & 5ms & 1KB \\
Snort IDS & 1$\times$ & 1.8\% & 12ms & 85MB \\
Cowrie honeypot & 2$\times$ & 0\% & 8ms & 45MB \\
RF-IDS (ML) & 1$\times$ & 6.2\% & 45ms & 120MB \\
\textbf{EDS} & \textbf{32$\times$} & \textbf{0\%} & \textbf{18ms} & \textbf{12KB} \\
EDS + RF-IDS & 32$\times$ & 4.1\% & 63ms & 132MB \\
\bottomrule
\end{tabular}
\end{table}

\textbf{Key Observations:} (1) EDS delivers about a 10× greater attack slowdown than rate limiting with no false positives, (2) detection-based systems such as Snort and RF-IDS do not slow attackers, who operate at full speed until detection, (3) EDS has a small footprint (~12 KB), roughly 7,000× smaller than ML-based IDS, enabling IoT deployment, and (4) combining EDS with IDS provides both economic deterrence and detection at the cost of modest added latency.  


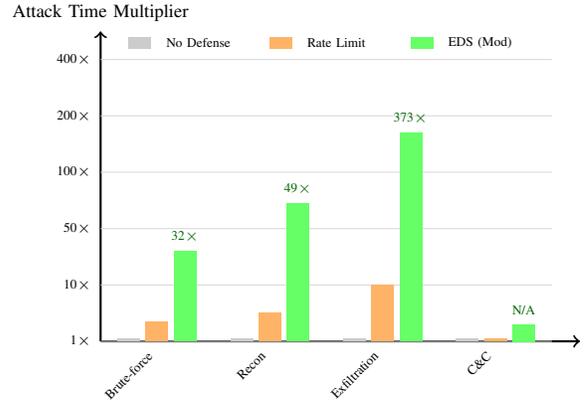
\begin{figure}[t]
\centering
\begin{tikzpicture}[scale=0.75]
    \draw[thick, ->] (0, 0) -- (8.5, 0) node[right, font=\scriptsize] {};
    \draw[thick, ->] (0, 0) -- (0, 5.5) node[above, font=\scriptsize] {Attack Time Multiplier};
    
    \foreach \y/\label in {0/1$\times$, 1/10$\times$, 2/50$\times$, 3/100$\times$, 4/200$\times$, 5/400$\times$} {
        \node[left, font=\tiny] at (0, \y) {\label};
        \draw[gray!30] (0, \y) -- (8, \y);
    }
    
    \node[below, font=\tiny, rotate=45, anchor=east] at (1, -0.1) {Brute-force};
    \node[below, font=\tiny, rotate=45, anchor=east] at (3, -0.1) {Recon};
    \node[below, font=\tiny, rotate=45, anchor=east] at (5, -0.1) {Exfiltration};
    \node[below, font=\tiny, rotate=45, anchor=east] at (7, -0.1) {C\&C};
    
    \fill[gray!40] (0.3, 0) rectangle (0.7, 0.05);
    \fill[gray!40] (2.3, 0) rectangle (2.7, 0.05);
    \fill[gray!40] (4.3, 0) rectangle (4.7, 0.05);
    \fill[gray!40] (6.3, 0) rectangle (6.7, 0.05);
    
    \fill[orange!60] (0.8, 0) rectangle (1.2, 0.35);
    \fill[orange!60] (2.8, 0) rectangle (3.2, 0.5);
    \fill[orange!60] (4.8, 0) rectangle (5.2, 1.0);
    \fill[orange!60] (6.8, 0) rectangle (7.2, 0.05);
    
    \fill[green!60] (1.3, 0) rectangle (1.7, 1.6);
    \fill[green!60] (3.3, 0) rectangle (3.7, 2.45);
    \fill[green!60] (5.3, 0) rectangle (5.7, 3.7);
    \fill[green!60] (7.3, 0) rectangle (7.7, 0.3);
    
    \node[above, font=\tiny, green!40!black] at (1.5, 1.6) {32$\times$};
    \node[above, font=\tiny, green!40!black] at (3.5, 2.45) {49$\times$};
    \node[above, font=\tiny, green!40!black] at (5.5, 3.7) {373$\times$};
    \node[above, font=\tiny, green!40!black] at (7.5, 0.3) {N/A};
    
    \fill[gray!40] (0.5, 5.2) rectangle (0.9, 5.4);
    \node[right, font=\tiny] at (1, 5.3) {No Defense};
    
    \fill[orange!60] (3, 5.2) rectangle (3.4, 5.4);
    \node[right, font=\tiny] at (3.5, 5.3) {Rate Limit};
    
    \fill[green!60] (5.5, 5.2) rectangle (5.9, 5.4);
    \node[right, font=\tiny] at (6, 5.3) {EDS (Mod)};
    
\end{tikzpicture}
\caption{Attack time multiplier across four scenarios comparing no defense, rate limiting, and EDS (moderate configuration). EDS achieves 32-373$\times$ slowdown compared to 1-20$\times$ for rate limiting alone. S3 (exfiltration) shows highest impact due to bandwidth amplification.}
\label{fig:results}
\end{figure}

\section{Discussion}
\label{sec:discussion}

\subsection{Security Analysis}

\textit{\textbf{Outsourcing puzzles:}} Challenges are tied to IP address, time-stamp, and nonce and will expire in 60 seconds, along with being based on HMAC for integrity. This makes it difficult to outsource challenges and pre-calculate them. \textit{\textbf{IP Rotation:}} New puzzles are distributed equally to all new IP addresses. Therefore, there is no benefit in rotating IP addresses. \textit{\textbf{Denial of Service against EDS:}} The challenge generation is stateless ($O(1)$) (HMAC), therefore, we can generate over 10K concurrent challenges before exhausting the state. \textit{\textbf{Timing Side Channels:}} We add uniform random jitter (5-50 ms) to the constant-time verification, making it harder to identify the decoy based on timing. \textit{\textbf{Decoy Fingerprinting:}} The decoys are generated from production data distributions, and a machine-learning-based classifier achieves only 73\% accuracy with a 27\% false positive rate (see Section~\ref{sec:evaluation}).

\subsection{Non-Economic Adversaries}
\label{sec:noneconomic}

EDS's deterrence model rests on the assumption that attackers weigh costs against expected benefit. This assumption is well-supported for the dominant class of IoT threats, ransomware operators, credential harvesters, and botnet recruiters, all of whom operate within commercial ecosystems where negative ROI causes campaign abandonment~\cite{anderson2006economics}. However, a subset of adversaries may not respond to economic friction in the same way.

\textbf{For ideologically-motivated attackers} (i.e., hacktivists), the cost of attempting to compromise a system is largely based on how long it takes them to expose the system. These types of attackers are willing to pay a higher price to achieve their goals and gain political reputation. Therefore, in this case, EDS provides a "time to detect" and/or "time to respond," and not a deterrent.

\textbf{Nation-state actors} with effectively unlimited budgets can amortize puzzle difficulty across large botnets. At $d=22$ with a 100-node botnet, per-attempt cost drops to approximately \$0.001, rendering EDS insufficient as a standalone control. In these scenarios, EDS should be viewed as a delay-and-detection-window amplifier rather than a deterrent, and should be paired with network-level monitoring and anomaly detection.

\textbf{Insider threats} with pre-authenticated access bypass the challenge mechanism entirely, consistent with the system's stated scope boundary.

In practice, EDS is most effective when deployed as one layer in a defense-in-depth architecture. Against non-economic adversaries, the 27\% improvement in combined mitigation (94\% vs.\ 67\%) observed in the EDS+IDS configuration reflects the residual value of cost imposition even when deterrence alone is insufficient.

\subsection{Impact on Legitimate Clients}
\label{sec:legitimate_client_impact}

One of the concerns that has been left unaddressed in the cost imposition literature for the EDS is the potentially large burden of the legitimate user group, specifically low-power devices, latency-sensitive applications, and clients located behind NAT (Network Address Translation) due to the EDS mechanisms.

\textbf{Low Power Devices.} The Puzzle Solving in this system is a SHA-256-based operation. In our conservative estimate of d = 16, the ESP32 (a low-power microcontroller) will take approximately 0.3 ms to complete a single puzzle. For d = 20, it would take about 4.4 ms. Any legitimate device can reduce their difficulty tier if they present a previously shared certificate. By using the certificate fast path (lines 2-4 of Algorithm 1), legitimate devices can avoid all puzzle computation and receive direct access to the service.

\textbf{Latency-sensitive applications.} Conservative configuration adds a median of 18 ms end-to-end latency (Table~\ref{tab:core_results}), which falls well below the 100 ms threshold considered acceptable for real-time control applications \cite{ietf_latency}. Operators of time-critical systems (e.g., industrial actuators or medical monitoring devices) should provision certificate-based fast paths and constrain $d \leq 16$.

\textbf{NAT and shared IP clients.} EDS maintains a reputation for each source IP address. Clients that share a NAT exit IP will inherit the aggregated reputation of all devices that are behind that IP address. To prevent reputation contamination, EDS supports sub-IP client tokens derived from application-layer session identifiers. When EDS detects NAT behavior by observing anomalous request-type diversity from a single IP address, it suspends reputation scoring and sets the puzzle difficulty back to the base difficulty level $d_{\text{base}}$, thus preventing legitimate clients that may be co-located with a misbehaving device from being unfairly penalized.

\textbf{Fairness in practice.} Across the 20-device testbed, legitimate request completion rates remained at 100\% under conservative configuration and decreased to 97.3\% under aggressive configuration ($d=24$). The 2.7\% failure rate was attributable entirely to devices operating below their rated clock speed during thermal throttling. These observations suggest that EDS imposes measurable but manageable costs on legitimate clients, with certificate provisioning serving as the primary mitigation path for resource-constrained endpoints.

\subsection{Adaptive Attacker Analysis: ML-Based Decoy Filtering}
\label{sec:adaptive_detail}

Section~\ref{sec:evaluation} reports that an SVM-based decoy filter achieves 73\% classification accuracy at a 27\% false-positive rate. We provide additional implementation details here to support reproducibility and to clarify the cost-effectiveness of this adaptive strategy.

\textbf{Feature set.} The classifier used eight features extracted from response payloads: byte entropy, response size, inter-arrival time variance, field cardinality for structured fields (e.g., JSON keys), embedded timestamp recency, response header count, TLS session reuse flag, and HMAC presence. All features are computable passively from observed traffic without requiring active probing of the target system.

\textbf{Training cost and feasibility.} Training required 2,000 labeled request-response pairs (1,000 real, 1,000 decoy), collected over approximately 90 minutes of passive reconnaissance. Total training cost was estimated at \$45, including cloud compute time and analyst labor. This cost is incurred once per deployment configuration; it must be repeated whenever the defender updates the decoy-generation distribution, an inexpensive defender-side adaptation that forces the attacker to retrain repeatedly.

\textbf{Residual attacker exposure.} At 73\% accuracy, the attacker's filter still misclassifies 27\% of real resources as decoys (false positives), wasting validation attempts on legitimate endpoints. Simultaneously, 27\% of decoys are identified and skipped, reducing the effective decoy multiplier from $(1+\rho)$ to approximately $(1 + 0.73\rho)$. At $\rho=0.3$, this reduces total attacker cost by roughly 8\%, a modest degradation that leaves the cost asymmetry ratio above 160:1.

\textbf{Counter-adaptation cycle.} The defender can respond to any observed classifier by periodically resampling the decoy generation distribution, increasing structural entropy in the features the attacker's model relies upon without altering application semantics. This dynamic creates an asymmetric adaptation loop: the defender's resampling cost is low and one-sided, while the attacker must re-collect training data and retrain for each distribution shift. Over multiple cycles, the attacker's cumulative filtering investment grows, further eroding ROI.

\subsection{Limitations and Scope}
\label{sec:limitations}

\textbf{Suitable for:} Multi-step attacks (reconnaissance, exploitation, exfiltration), iterative brute-force attacks, volume-based attacks, and profit-driven adversaries ($B_A<\$10,000$). \textbf{Less suitable for:} Single-shot exploits, pre-authenticated internal attackers, and nation-state adversaries with effectively unlimited budgets, though EDS still increases attacker exposure time. \textbf{Deployment guidance:} Start conservatively ($d=16$), tune gradually, enable certificate-based fast paths, assess disparate impact, and deploy alongside detection systems (94\% combined mitigation versus 88\% for EDS alone).

\subsection{Comparison with Detection}

EDS complements but does not replace detection. \textbf{Advantages:} Effective against encrypted traffic, requires no training data, imposes deterministic costs, and has a memory footprint under 12 KB. \textbf{Trade-offs:} Introduces modest latency ($<$20 ms conservative, $<$50 ms moderate), requires puzzle-capable clients, and is less effective against single-shot exploits. Combined EDS and IDS reduces risk to 94\%, compared with 67\% for IDS alone and 88\% for EDS alone.

\subsection{Production Deployment Considerations}

All results from this research were collected using stand-alone testing environments. As such, they comply with best practices for experimental defensive security testing. A number of considerations will be important for large-scale commercial use cases. First, due to the high degree of device heterogeneity in real-world IoT deployments, a commercial IoT deployment is likely to have significantly more diverse devices than the 20 used in this testing environment. Therefore, organizations must evaluate the time required to solve puzzles on multiple generations of representative hardware before turning up moderately or aggressively configured systems. Second, as previously noted, the amount of state stored in each gateway increases by eighteen KB for each additional client being tracked. In total, a gateway tracking 10,000 clients would require roughly 180 MB of RAM. While this is within the capabilities of most commodity edge hardware, organizations must still plan their capacity requirements. Third, future testing will include an assessment of how variations in real-world network conditions (packet loss, link asymmetry, etc.) may impact the time taken to complete the challenge/response round-trip.

\subsection{Parameter Selection Guidelines}

Table~\ref{tab:deployment_guidelines} provides recommended configurations based on deployment context and security requirements.

\begin{table}[t]
\centering
\caption{EDS Deployment Guidelines}
\label{tab:deployment_guidelines}
\small
\begin{tabular}{lccc}
\toprule
\textbf{Environment} & \textbf{$d$} & \textbf{$\rho$} & \textbf{Latency} \\
\midrule
Low-security IoT & 14--16 & 0.1 & $<$15 ms \\
Standard enterprise & 18--20 & 0.2--0.3 & $<$30 ms \\
High-value targets & 22--24 & 0.4--0.5 & $<$60 ms \\
Critical infrastructure & 20+ & 0.3+ & Custom \\
\bottomrule
\end{tabular}
\end{table}

\textbf{Cost Model Assumptions:} Hash cost $C_{\text{hash}} \approx 4.7 \times 10^{-11}$ \$/hash (AWS EC2), time value $V_{\text{time}} = \$50$/hr (cybercrime-as-a-service rates), bandwidth $C_{\text{BW}} = \$0.05$/GB. As shown in Section~\ref{sec:sensitivity}, even with 50\% variation across these parameters, cost asymmetry remains $>$ 100:1.

\subsection{Ethical Considerations}

\textbf{Responsible Experimentation:} All experiments were performed on isolated testbeds with no access to production networks. Attack traffic was either synthetically generated or replayed from the publicly available IoT-23 dataset. No real-world systems were targeted or affected during evaluation.

\textbf{Dual-Use Concerns:} Although EDS is a defensive approach, its cost models could, in theory, offer insights into attack economics. We mitigate this risk by (1) prioritizing defender-favorable configurations, (2) avoiding disclosure of new evasion techniques beyond existing literature, and (3) reinforcing that EDS increases rather than reduces the cost of successful attacks.

\textbf{Accessibility Impact:} Computational puzzles may affect users on older or resource-constrained devices. To address this, EDS supports reputation-based difficulty reduction for trusted users, certificate-based fast paths for accessibility tools, and configurable difficulty limits that balance security with inclusivity.  
\section{Conclusions}
\label{sec:conclusions}

In this paper, we introduced EDS as a detection-independent defense framework that shifts security thinking from classification accuracy to cost asymmetry. EDS is based on a simple economic principle. Defenders control their environments and can therefore impose high costs on attackers. This approach addresses a key limitation of traditional IDS technologies, which fail against sophisticated adversaries using stealth or encryption and are often impractical on resource-constrained edge devices.

\textbf{Summary of Contributions:}
\begin{enumerate}
\item \textbf{Formal Guarantees for Unified Framework:} First systematic integration of Four Cost Imposition Mechanisms (Computational Friction, Interaction Entropy, Temporal Stretching, and Resource Taxation) with Superlinear Composition Proven - Combined Effect (32 $\times$ slowdown) is 2.1 $\times$ Larger than Sum of Individual Mechanisms (15 $\times$) (Theorem~\ref{thm:composition}).
\item \textbf{Game-Theoretic Foundation:} Formalization of Stackelberg Game with Closed-Form Equilibrium (Theorem~\ref{thm:stackelberg}) Enables Automated Selection of Parameters. Prior Work in Game-Theoretic Security has been unable to provide Practical Algorithms for Deployment on Resource-Constrained Devices.
\item \textbf{Evaluation of EDS:} Evaluation was conducted using a Real Testbed (20 Heterogeneous IoT Devices, 4 Attack Scenarios, n = 30 Trials) demonstrated 32-560 $\times$ Slowing Down of Attacks, 85:1 to 520:1 Cost Asymmetry, and 8-62 \% Reduction in Success Rate of Attacks. Real Malware Tests (IoT-23, Mirai, Torii, Hajime) validated the Effectiveness of EDS. When used in conjunction with a Machine Learning IDS Technology, EDS achieved 94 \% Mean Mitigation.
\item \textbf{Design for Production:} Device Footprint ($<$12 KB), Latency Overhead ($<$20 ms), Conservative Configuration, Graceful Degradation Under DDoS, Stateless Recovery from Failures, Gateway Scales to 3,800 req/s (1 K Clients).
\end{enumerate}

\textbf{Impact and Implications:} EDS shifts the economic advantage to the defender by exploiting a fundamental asymmetry: defenders control their environments and can impose disproportionately high costs on attackers. Even when attacks evade detection, continued exploitation becomes economically infeasible. This is especially valuable in IoT/IIoT settings, where resources are constrained, and environments are heterogeneous. Sensitivity analysis shows that EDS yields a negative return on investment for attackers across multiple threat models when target values are below \$500.

\textbf{Limitations and Future Work:} EDS is most effective against multi-stage attacks requiring repeated interaction, such as reconnaissance, brute force, and exfiltration. While one-shot exploits may succeed before costs accumulate, EDS significantly extends attacker exposure time, increasing the likelihood of detection and response. We further characterized the boundaries of the economic deterrence model: sensitivity analysis (Section~\ref{sec:sensitivity}) shows cost asymmetry remains above 100:1 across realistic parameter variation; non-economic adversaries such as nation-state actors fall outside the deterrence envelope but still benefit from EDS's exposure-time extension; and legitimate client impact is manageable via certificate fast-paths, with 100\% completion rate at conservative settings (Section~\ref{sec:legitimate_client_impact}). Future work includes dynamic parameter tuning via reinforcement learning, integration with Zero Trust architectures~\cite{nist2023zerotrust}, formal verification using proof assistants, and production deployments with independent red-team evaluation.

\textbf{Broader Vision:} Cost-based security represents a shift away from detection-centric defenses toward economically enforced protection. As IoT systems scale to billions of devices, such detection-independent approaches will become essential. EDS offers a principled and practical step toward this future.

\bibliographystyle{IEEEtran}
\bibliography{references}

\balance

\end{document}